\documentclass[prb,twocolumn,floats]{revtex4}

\usepackage{graphicx}

\begin{document}

\title {Quasiparticle vanishing driven by geometrical frustration}

\author{A. E. Trumper, C. J. Gazza, and L. O. Manuel}

\affiliation {Instituto de F\'{\i}sica Rosario (CONICET) and
Universidad Nacional de Rosario, Boulevard 27 de Febrero 210 bis,
(2000) Rosario, Argentina}

\vspace{4in}

\date{\today}

\begin{abstract}
We investigate the single hole dynamics in the triangular $t\!-\!J$ model.
We study the structure of the hole spectral function, assuming
the existence of  a 120$^{\circ}$ magnetic N\'eel order.
Within the self-consistent Born approximation (SCBA) there is a strong
momentum and $t$ sign dependence of the spectra, related to the underlying
magnetic structure and the particle-hole asymmetry of the model.
For positive $t$, and in the strong coupling regime, we find that the low
energy quasiparticle excitations vanish outside the neighbourhood of the
magnetic Goldstone modes; while for negative $t$ the quasiparticle
excitations are always well defined. In the latter, we  also find
resonances of magnetic origin whose energies scale as $(J/t)^{2/3}$ and can
be identified with string excitations.
We argue that this complex structure of the spectra is due to the subtle
interplay between magnon-assisted and  free hopping mechanisms.
Our predictions are supported by an excellent agreement between the SCBA
and the exact results  on finite size clusters.
We conclude that the conventional quasiparticle picture can be broken by the
effect of geometrical magnetic frustration.
\end{abstract}

\maketitle
\section{Introduction}
The study of doped antiferromagnets has become a central topic
in the area of strongly correlated electrons systems, mainly after Anderson's
proposition\cite{anderson87} about the probable existence of non-Fermi
liquid behavior once a Mott insulating state is hole-doped.
A valuable experiment that gives access
to single-particle properties of many-electron systems is the
angle resolved photoemission spectroscopy (ARPES). Even though in the last years there
has been a rapid pace
of developments in ARPES experiments, there are still controversies in many strongly
interacting cases, like the cuprates, where  the difficulty to distinguish a coherent
$\delta$-function peak  from an incoherent part  of the spectra does
not allow
a proper interpretation of the available data\cite{damascelli03,laughlin97}. A useful
microscopic model widely used to describe  the low-energy physics involved in
photoemission experiments of
antiferromagnetic Mott insulators\cite{wells95} is  the $t\!-\!J$ model on the
square lattice \cite{dagotto94}.

Attention has also been concentrated on triangular Mott insulators.
These systems are attractive due to the presence of both,  strong correlation and geometrical
magnetic frustration, leading to non-conventional behavior.
Experimental realizations of such
systems are $Cs_{2}CuCl_{4}$\cite{coldea01},
 the $\kappa$ family of the organic charge transfer salts
$(BEDT-TTF)_{2} X$ \cite{mckenzie97}, and the silicon surfaces K/Si (111):B and
SiC(0001)\cite{johansson96,manuel03}. In a recent publication, Koretsune and
Ogata\cite{koretsune02} have studied the effect of finite doping in the triangular Heisenberg model.
They found that a  resonating-valence-bond (RVB) ground state is favored
when $t$ is positive, while a Nagaoka's ferromagnetic ground state
is stabilized for negative $t$.
More recently, the role of the $t$ sign and frustration has been addressed in the
superconductor $Na_{x}CoO_2$ where the $Co$ atoms form the triangular lattice
\cite{ogata03}. These remarkable issues lead  us to investigate the spectral
function of a hole in a
triangular Mott insulator, so as to analize the consequences of the geometrical
frustration and the $t$ sign on the hole motion. In particular, our aim is to study the existence
of quasiparticle (QP) excitations
and  higher-energy features in the spectral functions. To this purpose,  we focus on the hole
dynamics in the triangular $t\!-\!J$ model,
and we solve it using the Lanczos exact diagonalization method  and the self-consistent Born
approximation (SCBA).
In the non-frustrated case, it is known that the SCBA  reproduces quite well the
exact diagonalization
results on small clusters\cite{martinez91} and the angle-resolved
photoemission spectroscopy (ARPES) experiments\cite{wells95}.
In particular, the SCBA
predicts the existence of a well defined quasiparticle peak (magnetic polaron) at low energies
for the whole
Brillouin zone (BZ). Furthermore, the SCBA is able to capture finite lifetime resonances above the
QP peaks. These resonances have been explained by the string picture since their energies
 scale as $(J/t)^{2/3}$, as it happens in the Ising limit.
In our triangular frustrated case, the presence of the $120^{\circ}$ N\'eel order generates two
mechanisms  for hole motion in the SCBA effective Hamiltonian, one being tight-binding
like (or free) and the other, magnon-assisted. We will show that the $t$ sign tunes the
 interference between both  mechanisms, producing
pronounced differences in the low energy as well as in the higher energy region of
the spectral functions.
For the strong coupling regime, we find an excellent agreement between
 the SCBA and the exact results  on  a 21 sites cluster.
 In the thermodynamic limit, the SCBA spectral functions present a rich structure.
 Notably, for $t\! < \!0$ a quasiparticle excitation is always defined  for all ${\bf k}$
 of the Brillouin zone, while for $t\! >\!0$  in
the strong coupling regime the quasiparticle signal survives only around the Goldstone
magnetic wave vectors. By computing the quasiparticle energy scaling with $J/t$ it is possible
to get some insight of  the magnetic cloud surrounding the hole\cite{didier}. 
We argue that
for $t<0$ the local environment around the hole enhances
its AF character, while for $t>0$ the ferromagnetic character is favored. 
%*********************** MODIFICADO *****************************
We find that the single-hole dynamics in the thermodynamic limit behaves in the opposite
way as in very small clusters, like the three-sites problem\cite{koretsune02}.
%***************************************************************

Furthermore, for small values of $J/t$ and $t\!<\!0$, higher energy resonances
above the quasiparticle peak appear showing a $(J/t)^{2/3}$ energy scaling.  Although the
triangular $t-J^{z}$ model is completely different to the square lattice case, we continue
identifying these resonances with  string-like excitations.
 For $t\!>\!0$, instead, we do not observe higher energy resonances  that can be identified
 with strings. This behavior observed at higher energies can be also traced back to
 the particular features of the magnetic environment around the hole mentioned above.

The paper is organized as follows. In section II we obtain the effective Hamiltonian and the
one hole properties within the SCBA. In section III we present the results:
(IIIA) the comparison
between  exact and  SCBA results in finite size clusters, (IIIB) the quasiparticle
features, and (IIIC) the general description of the spectral functions.
Finally, in section IV we draw the concluding remarks.

\section{ t-J model and self consistent Born approximation}
The  $t\!-\!J$ model is defined as
\begin{equation}
{H}=-t\sum_{\langle {i},{j} \rangle }(\hat{c}^{\dagger}_{i,\sigma}
                                         \hat{c}_{j,\sigma}+ h.c.)+
  J\sum_{\langle {i},{j} \rangle }
{{\bf {S}}}_{i} \cdot {{\bf {S}}}_{j}~,
\label{tj}
\end{equation}

\noindent where the first term represents the hopping between nearest
neighbors ($<i,j>$) of the triangular lattice,  with the constraint of no double
occupancy, $\hat{c}_{i,\sigma}=c_{i,\sigma} (1-n_{i,-\sigma}$),
and the second term represents the antiferromagnetic (AF) Heisenberg part with $J\!>\!0$.

The study of the single hole motion in an AF  requires a correct
description of the magnetic low energy sector since the injected
hole will couple strongly with these spin excitations.
In the literature there is firm numerical \cite{bernu92}
and analytical \cite{manuel98} evidence that the triangular Heisenberg
antiferromagnet has a $120^ {\circ} $ N\'eel order ground state, and
that spin wave theory reproduces quite well its low energy
behavior\cite{trumper00}. So that, we assume an ordered  ground state
with a magnetic wave vector ${\bf Q}=(4\pi/3,0)$ lying in the
$x\!-\!z$ plane, and spin waves as the low energy excitations.
In order to simplify the calculation we perform a unitary transformation
to local spin quantization axis --primed operators-- in order
 to have a ferromagnetic ground state in the  $z^{\prime}$ direction.
Then, we use the spinless fermion representation for
electrons
$\hat{c}^{\prime}_{i \uparrow}=h^{\dagger}_i, 
\hat{c}^{\dagger \prime}_{i \downarrow}=h_i S^{-}_i,$
and the Holstein-Primakov representation for spin
operators (to order $1/S$)
$S^{x^\prime}_i \sim \frac{1}{2}(a^{\dagger}_i+a_i), 
S^{y^\prime}_i \sim \frac{{\it i}}{2}(a^{\dagger}_i-a_i), 
S^{z^\prime}_{i} = \frac{1}{2}-a^{\dagger}_i a_i$,
where $h^{\dagger}_i$ creates a spinless hole at site $i$, and $a^{\dagger}_i$
creates a spin deviation. 

%***************MODIFICADO**************************
We replace the above expresions in eq. (\ref{tj}) and,   
after Fourier and Bogoliubov transformations, the effective Hamiltonian results
\begin{eqnarray}
\label{HSW}
H  =  \sum_{\bf k} &  \epsilon_{\bf k}h^{\dagger}_{\bf{k} } h_{\bf{k} } &
 + \sum_{\bf q}  \omega_{\bf q} \alpha^{\dagger}_{\bf q} \alpha_{\bf q}
\nonumber  \\
&  - t\sqrt{\frac{3}{N}} & \sum_{\bf k, q}
\left[M_{\bf kq}h^{\dagger}_{\bf k}
h_{\bf k-q}\alpha_{\bf q} + h.c. \right]
\end{eqnarray}

\noindent with $\epsilon_{\bf k}=-t\gamma_{\bf k}$ and
$
\omega_{\bf q}=\frac{3}{2}J\sqrt{(1-3\gamma_{\bf q})(1+6\gamma_{\bf q})}, 
$
the bare hole and magnon dispersions, respectively.
$$
M_{\bf kq}=i (\beta_{\bf k}v_{-\bf{q}} -\beta_{\bf k-q}u_{\bf q}) \nonumber
$$
is the bare hole-magnon vertex interaction with the geometric factors
$\gamma_{\bf k}=\frac{1}{3}\sum_{\bf e}\cos({\bf k}.{\bf e})$ and $
\beta_{\bf k}=\sum_{\bf e}\sin({\bf k}.{\bf e})$
(${\bf e}$'s are the positive vectors to nearest neighbors), and
$u_{\bf q}$ and $v_{\bf q}$ are the usual Boguliubov coefficients.
%**********************************************

\noindent In the Hamiltonian (\ref{HSW}) --constant terms related to
magnetic energy are omitted--
the free hopping hole term implies a finite probability of the hole to move without
emission or absorption of magnons. This is a direct consequence of the underlying
{\it non-collinear} magnetic structure.
The hole-magnon interaction adds another mechanism for the charge carrier motion
which is magnon-assisted.
The latter is responsible for the spin-polaron formation when a hole is
injected in a non-frustrated antiferromagnet\cite{kane89,martinez91}.
We will show the existence of a subtle interference between both
processes that turns out to be dependent on the $t$ sign.
An important quantity that allows us to study the interplay between such
processes is the retarded hole Green function that is defined as
$G_{\bf k}^{h}(\omega)= \langle AF| h_{\bf k}\frac{1}{(\omega+{\it i}\eta^{+}-
H )}
h^{\dagger}_{\bf k} |AF\rangle$, where $|AF\rangle$ is the undoped antiferromagnetic
ground state with a $120^{\circ}$ N\'eel order.
In the SCBA $|AF\rangle$  is the spin wave AF ground state, and after neglection of the
crossing diagrams, the Dyson's equation
for the self energy at
zero temperature results
$$
\Sigma_{\bf k}(\omega )=\frac{3t^2}{N_s}\sum_{\bf q}
\frac{\mid M_{\bf kq}\mid^{2}}{\omega -\omega_{\bf q}-\epsilon_{{\bf k}-{\bf q}}
-\Sigma_{\bf k-q}(\omega-\omega_{\bf q})}.
$$
We have solved numerically this self-consistent equation for
$\Sigma_{\bf k}(\omega)$, and calculated the hole spectral function
$
A_{\bf k}(\omega ) = - \frac{1}{\pi}ImG_{\bf k}^{h}(\omega),$
and the quasiparticle (QP) weight
$
z_{\bf k}=\left(1-
\frac{\partial \Sigma_{\bf k}(\omega)}{\partial \omega}\right)^{-1}\vert_{E_{\bf k}},$
where the QP energy is given by $E_{\bf k}=\Sigma_{\bf k}(E_{\bf k})$.

\section{Results}
\subsection{Exact and SCBA comparison}
\begin{figure}[ht]
\vspace*{0.cm}
\includegraphics*[width=0.45\textwidth]{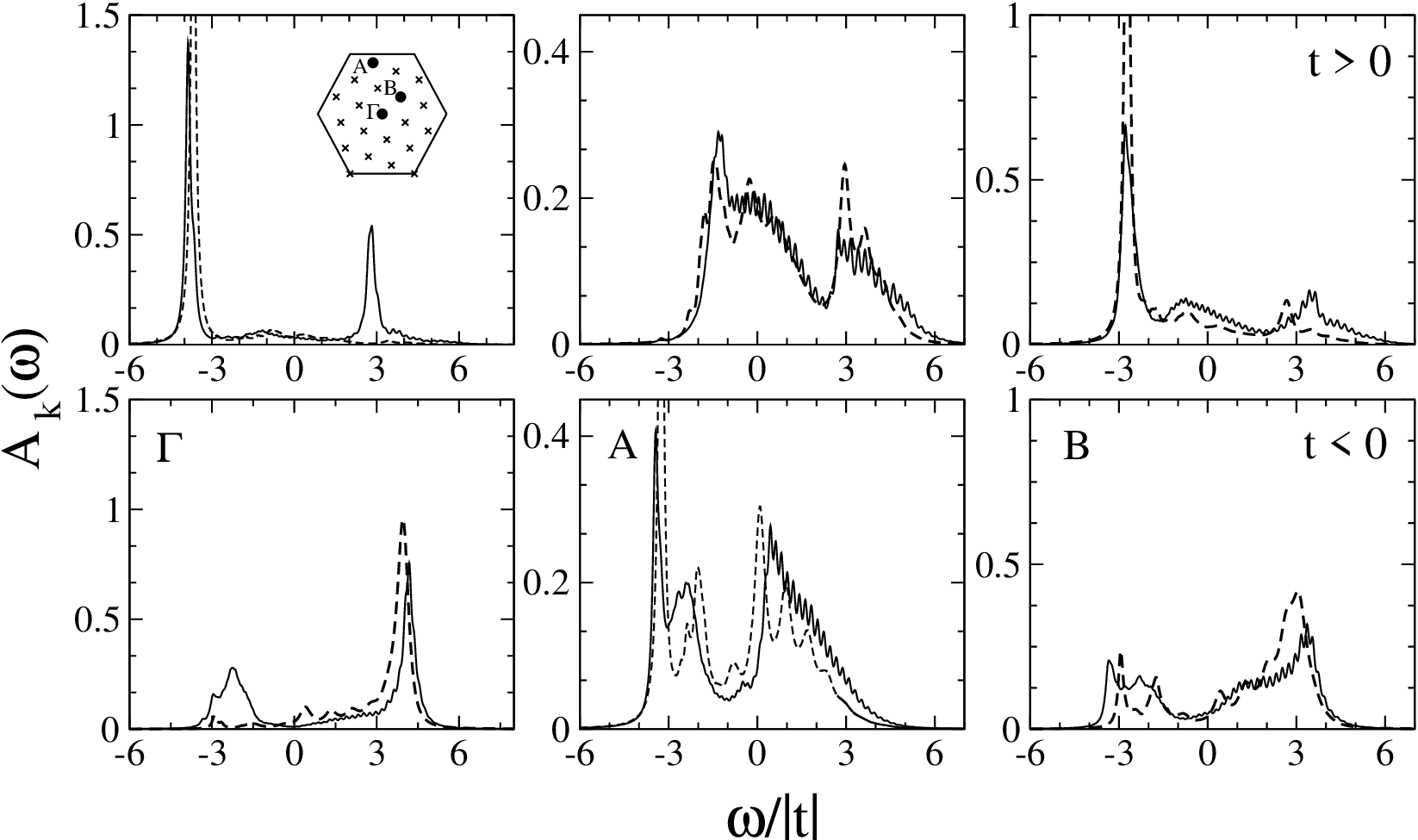}
\caption{Spectral functions versus frequency for  $J/|t|=0.4$ and
$N=21$ for momenta $\Gamma=(0,0),A=\frac{4\pi}{N}(-1,3\sqrt{3})$ and
$B=\frac{4\pi}{N}(2,\sqrt{3})$.  ${\bf k}$ shown as filled circles
in the inset (the crosses represent the other momenta).
The upper (lower) panel displays the results for positive (negative)
$t$. The solid and dashed lines are the exact and SCBA results,
respectively. $\eta=0.1$ and $\omega=10000$ has been used.}
\label{fig1}
\end{figure}

We have computed the one hole spectral function $A_{\bf k}(\omega )$ using the
Lanczos exact diagonalization method (ED)
and the SCBA on finite size clusters. To test the reliability of the SCBA
predictions we  have compared both techniques on the
$ N=21$ sites cluster,  which
preserves  the spatial symmetries of the infinite triangular
lattice\cite{bernu92}.
 In general, we have found a good agreement between  ED and  SCBA  results for all momenta
of the BZ and for $J/|t| \lesssim 1$.
In Fig. \ref{fig1} we show the spectral functions at three representative
$\bf k$ points of the BZ (see inset Fig. \ref{fig1}) in the strong coupling
regime, $J/|t| =0.4$, and for both signs of the transfer integral $t$.

The exact structure
of the spectral functions strongly depends on the momentum because the hole strongly
scatter off  antiferromagnetic fluctuations\cite{dagotto94}. Furthermore, the geometrical
frustration is manifested in a marked dependence of the spectral
function with the $t$ sign. Remarkably, the SCBA reproduces very well
these exact results in the whole range of energies.
In some cases from the scale of the
figure it cannot be clearly distinguished a quasiparticle peak at low energies.
By decreasing $\eta^+$ appropriately we have checked, however,
 the existence of a QP peak for all  $\bf k$ of the BZ. In particular,
for negative (positive) $t$ the $ \bf k=A$ ($\bf k=\Gamma$)  point becomes the
momentum of the quasiparticle ground state.
The energy of the hole
can be splitted in two contributions: $E^{h}\equiv E_{N-1}-E^{gs}_{N}
=E^{h}_{loc}+E^{h}_{deloc}$,
where $E^{h}_{loc}=E^{t=0}_{N-1}-E^{gs}_N$ is the energy of the localized hole,
and $ E^{h}_{deloc}=E_{N-1}-E^{t=0}_{N-1}$ is the energy gain due to
delocalization of the hole\cite{everts98}.
As the SCBA does not take into account the localized hole energy
we have shifted the exact energy spectra by an amount
$E^{h}_{loc}=-0.5686 J$ for $N=21$
in order to make a proper comparison with the SCBA results.
The agreement between both techniques is similar to that found in
the $4\times4$ square lattice besides the presence of stronger
antiferromagnetic fluctuations in our frustrated case.
It was shown that this agreement is a consequence of the small
leading nonzero contribution to the self energy from the
crossing diagrams\cite{liu92}.

\subsection{Quasiparticle}

\begin{figure}[ht]
\vspace*{0.cm}
\includegraphics*[width=0.45\textwidth]{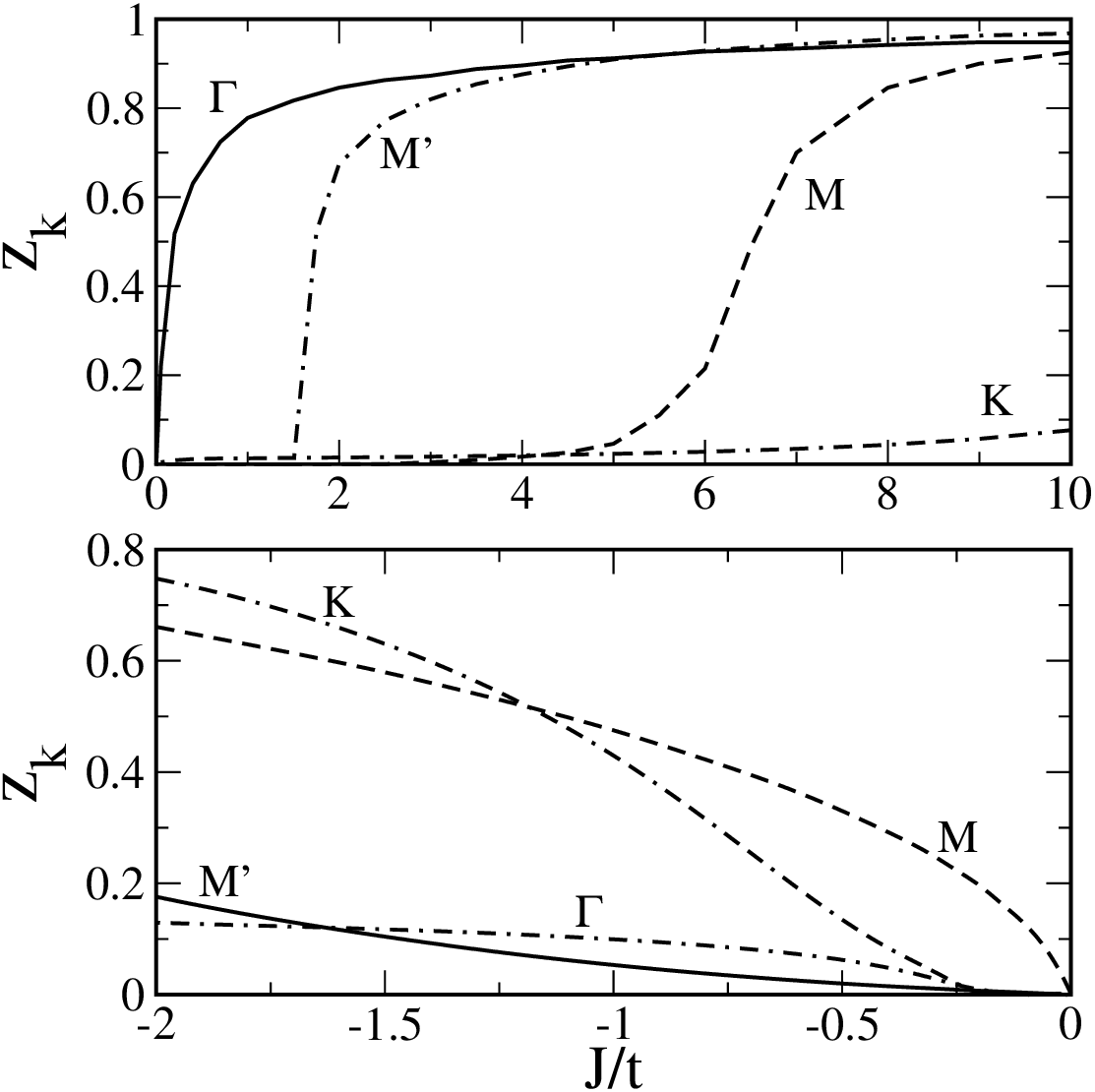}
\caption{Quasiparticle weight vs $J/t$. Upper (lower) panel is for $t>0$  ($t<0$).
 The location of the $\Gamma, M^{\prime}, K, M$ is displayed in Fig. \ref{fig3} }
\label{fig2}
\end{figure}

The momentum and the $t$ sign dependence of the spectra aforementioned is
strengthened in the thermodynamic limit. In particular, we observe
important differences in the quasiparticle weight $z_{\bf k}$
through the BZ under the change of $t$ sign.
Using
lattice sizes up to $N=2700$
we have extrapolated the results to the thermodynamic limit.
 In Fig.\ref{fig2}  we  show the values
of $z_{\bf k}$  as a function of $J/t$ for representative points along high symmetry axes of
the BZ .
The most salient feature we have obtained is the vanishing of the quasiparticle weight
at some momenta for positive $t$. At the $M$ and $M^{\prime}$ points $z_{\bf k}=0$
below $J/t\sim2.5$
and $1.5$, respectively. It is worth noticing the robustness and then the rapid decay to
zero of $z_{M^{\prime}}$ as $J/t\rightarrow1.5$. A similar behavior has the QP signal
at $\Gamma$,
the ground state momentum, but it goes to zero only when $J/t \rightarrow 0$. Finally, the QP
signal is very weak, but finite, at the magnetic wave vector $K$ for all the $J/t$ range studied;
for instance, $z_{\bf k}\sim 0.008$ when $J/t=10$. Surprisingly, for  $J >> t$, $z_{ K}$ does not
seem to approach one as it turns out in the square lattice case\cite{martinez91} .
The non existence of quasi-particle excitations is a striking manifestation of the strong
interference between the free and magnon-assisted  hopping processes, tuned by the $t$ sign.
For the strong coupling regime and a large
region of the BZ, the coherence of the QP is lost because of the quite different
group velocities of the spin fluctuation and the tight binding hole motion, suggesting the
likelihood of a spin-charge separation scenario\cite{anderson87,laughlin97,beran96}
for the triangular $t-J$ model.

\begin{figure}[ht]
\vspace*{0.cm}
\includegraphics*[width=0.25\textwidth]{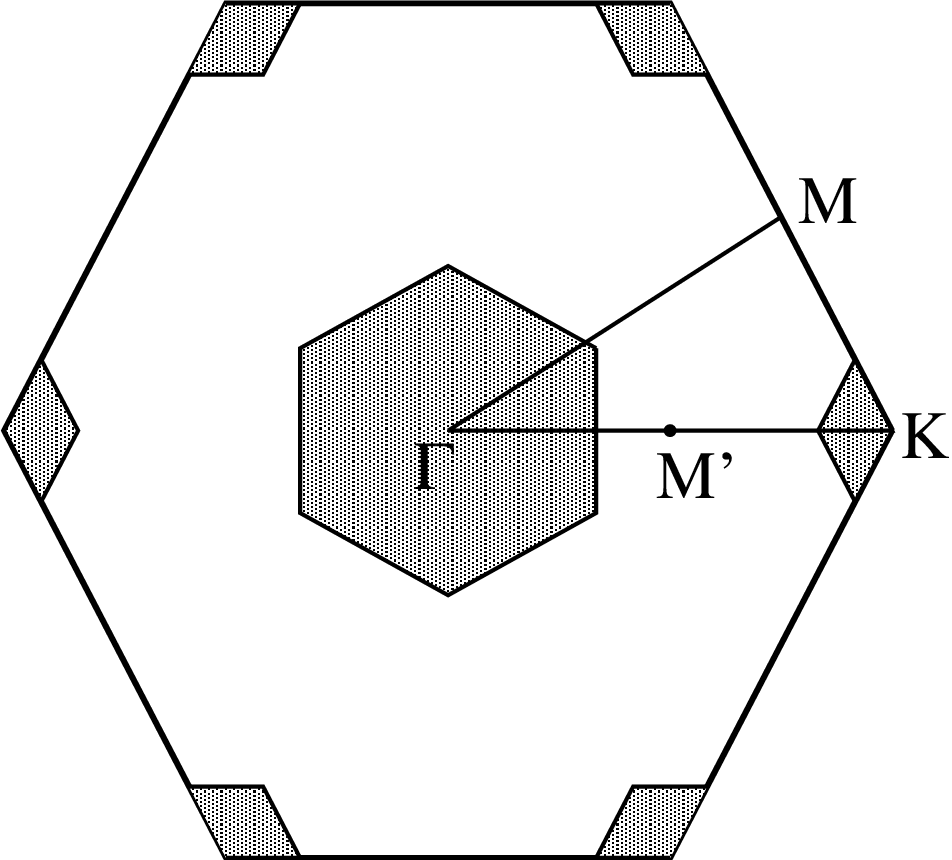}
\caption{
Triangular Brillouin zone. The shaded areas correspond to the regions where the
 QP weight is finite for $J/|t|=0.4$ and positive $t$.}
\label{fig3}
\end{figure}

On the other hand, for negative $t$  the quasiparticle weight is finite
for all momenta and for $J/t \neq 0$. This behavior is similar to the one
found in the non-frustrated case \cite{martinez91}. It is interesting to
note that now the QP ground state momentum is $M$ ($K$) for
$J/|t| \leq 1.2$ ($J/|t| > 1.2$).
The spectral weight of the QP becomes the most robust for the ground state momenta. The interchange in the ${\bf k}$ dependence of the
 QP weight, evidenced in a comparison
between both panels of Fig. \ref{fig2},
could be thought as a remnant of the particle-hole symmetry that shifts the momenta
by ${\bf Q}$ under $t$ sign reversal.

\begin{figure}[ht]
\vspace*{0.cm}
\includegraphics*[width=0.45\textwidth]{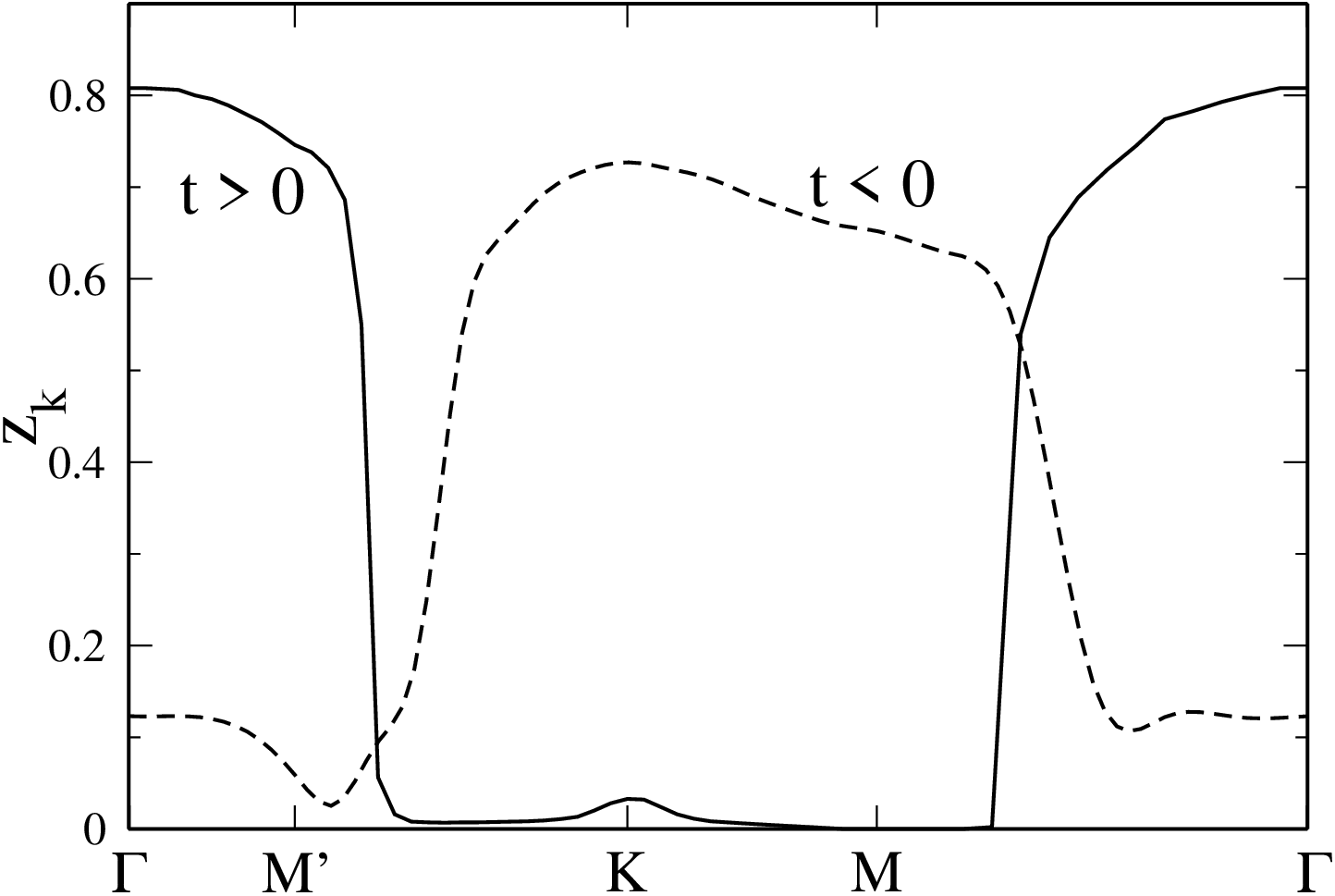}
\caption{Quasiparticle weight along the $\Gamma-M^{\prime}-K-M$ path
(see Fig. \ref{fig3})
Solid line is for positive $t$  and dashed line for negative $t$ . }
\label{fig4}
\end{figure}

In Fig. \ref{fig3} the shaded areas correspond to the regions where the QP weight is finite
for $J/|t|=0.4$ and positive $t$. There is no QP signal for momenta outside the neighborhood
of the magnetic Goldstone modes (${\bf k}={\bf 0}, {\bf Q}$).
The presence of QP signal in the non-equivalent vertices of the BZ is an artifact of the
used approximation: the spin wave magnetic ground state has a definite chirality
($\bf Q$), but our vertex interaction does not discriminate between
$\pm {\bf Q}$, resulting the whole spectral function degenerate
in both momenta.The shaded areas share
the same symmetries with the triangular magnetic Brillouin zone. We have noted that as
$J/t\rightarrow 0$ they shrink to the magnetic Goldstones modes. Conversely,
when $J/t \ge 2.5$ there is a finite QP weight anywhere and therefore, the full BZ becomes
shaded. For the particular value $J/|t|=2$, the QP weight vanishes only around the
$M$ point, as it
 is displayed by the solid line of Fig. \ref{fig4}. This figure also shows the strong
 ${\bf k}$ dependence of the QP weight for both $t$ signs. Again, a signature of the
  remnant particle-hole symmetry mentioned above is evidenced: the dashed line ($t<0$)
   resembles the solid one ($t>0$) shifted by the wave vector ${\bf Q}$.
In a previous work the hole dynamics in a triangular antiferromagnet
has already been studied using the SCBA \cite{azzouz96}, but the vanishing
of the quasiparticle weight for positive $t$ has been overlooked.
Our findings, however, are supported by a detailed and careful
treatment of the parameters involved: the convergence factor $\eta$,
the extrapolation to the
thermodynamic limit of the cluster sizes $N$, and the number of frequencies
$\omega$ \cite{liu92,tesislema}.

\begin{figure}[ht]
\vspace*{0.cm}
\includegraphics*[width=0.45\textwidth]{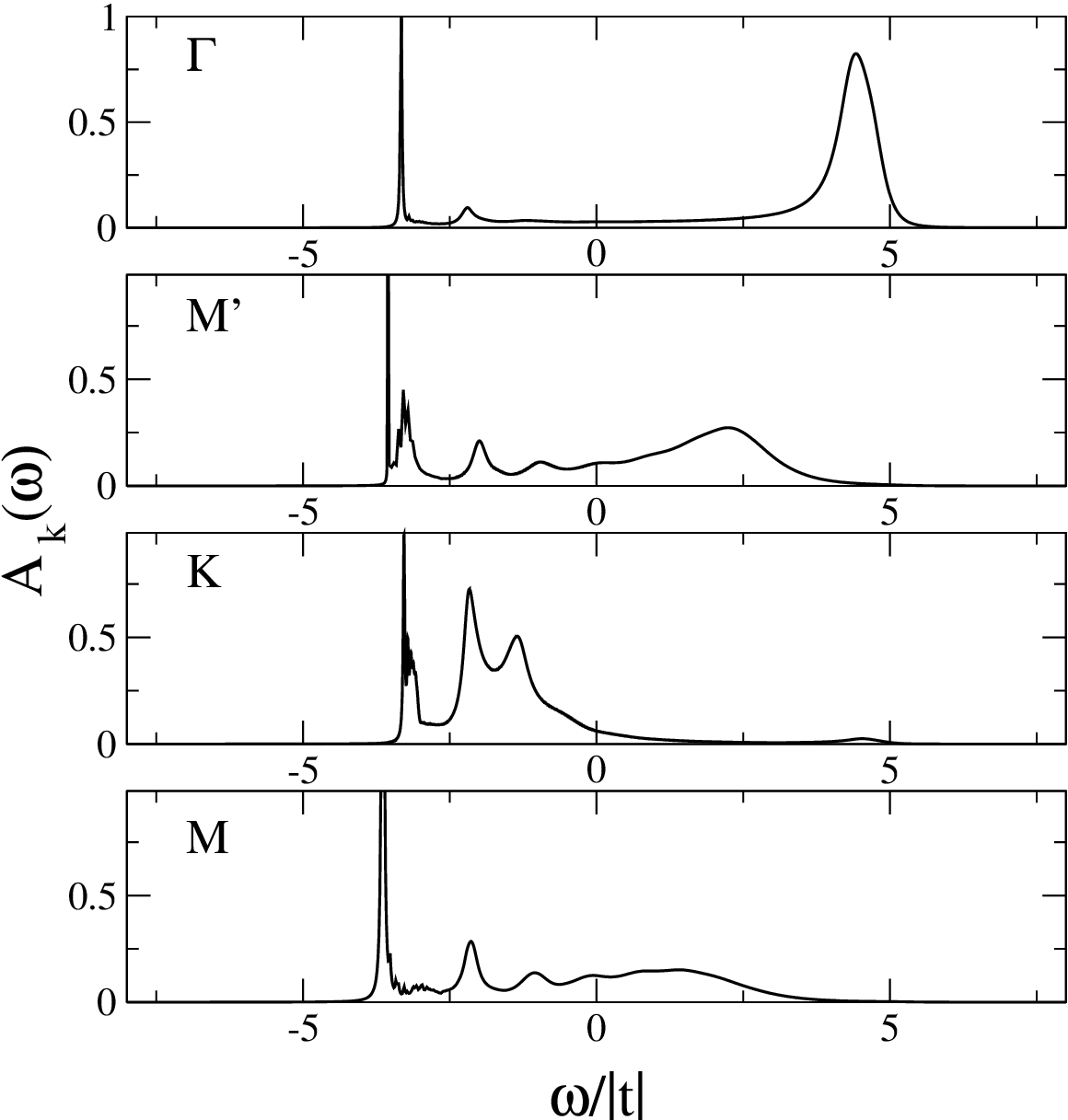}
\caption{Spectral function for the $\Gamma,M^{\prime},K,M$ points of the Brilluoin zone
for $J/|t|=0.4$ and negative $t$.}
\label{fig5}
\end{figure}

By computing the quasiparticle energy scaling with $J/t$ it is possible
to get some insight of  the magnetic cloud surrounding the hole\cite{didier}. It is known
that for a ferromagnetic polaron the QP energy scales as $\epsilon\sim(J/t)^{0.5}$, while for
an AF polaron it  follows a $\epsilon\sim(J/t)^{0.66}$ dependence. In our case we have
estimated a $\epsilon\sim(J/t)^{\alpha}$ scaling, being $\alpha\sim0.55$ ($\sim0.65$) for positive
(negative) $t$ in the range $ J/|t|\lesssim 0.1$.
A possible interpretation of  such results is the following: 
the non-collinear short range
magnetic correlations around the hole can be thought as a
superposition of ferro and AF correlations, and then
for $t<0$ the local environment around the hole enhances
its AF character, while for $t>0$ the ferromagnetic character is favored.
%**************** MODIFICADO ***********************************
It should be noticed that the opposite behavior holds in  a simple toy model
consisting of a three-sites $t-J$ model with one hole\cite{koretsune02}.
Actually, we have checked that for cluster sizes smaller than $N=21$, this behavior is
reversed. For instance, for $N=12$, $\alpha \sim 0.57$ for positive $t$ while for
negative $t$ it turns out $\alpha \sim 0.44$.
%*****************************************************************

\subsection{Spectral function structure}
\begin{figure}
\includegraphics*[width=0.45\textwidth]{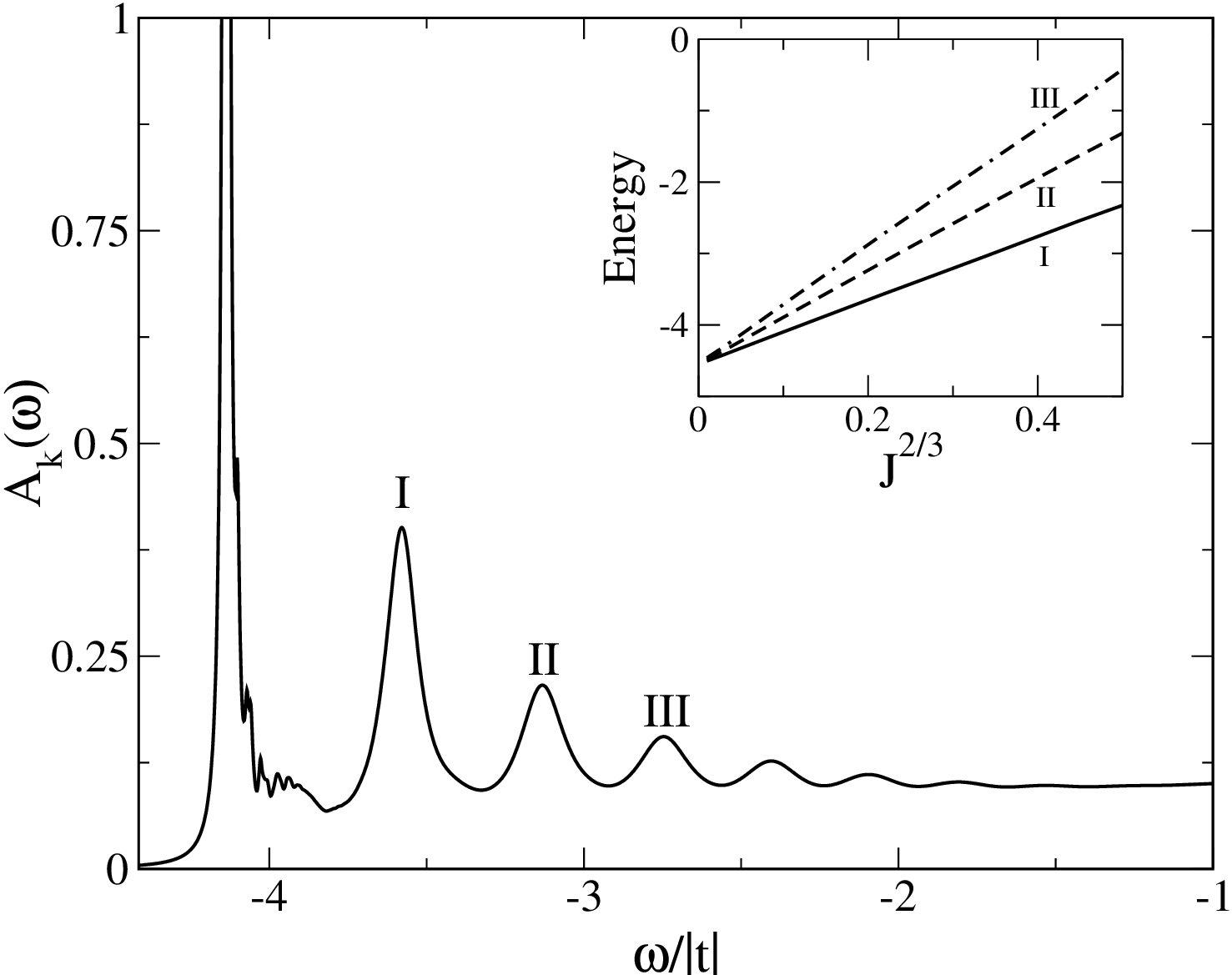}
\caption{ Spectral functions at the $M$ point for $J/|t|=0.1$ and negative $t$.
Inset: Energy scaling versus $(J/t)^{2/3}$ of the first three peaks labeled as I, II and III.}
\label{fig6}
\end{figure}

In this section we discuss the main features of the spectral functions in the
thermodynamic limit.
The arising structures can be traced back to the two hole motion processes
present in the effective Hamiltonian (\ref{HSW}).
The spectral functions at the $\Gamma$, $M^{\prime}$, $K$ and $M$
points are displayed in Fig. \ref{fig5}  for $J/|t|=0.4$ and negative $t$ .
The spectra extend over a frequency range of order $9t$; that is, the
non-interacting electronic energy scale. It is possible to discern
two distinct regions: a low energy sector of magnetic origin
(magnon-assisted), and a high energy sector with a broad resonance related
to the free hopping process. This resonance, located at an energy near
$\epsilon_{\bf k}$, has a finite lifetime of order $\sim 4J$ and a
dispersion of order $t$ .
Furthermore, the low energy sector  consists of a
quasiparticle peak and higher energy resonances with a narrow dispersion of order $J$
as the momentun varies. We will show below that the latter resonances can be identified with
string excitations. At $\Gamma$ and $M^{\prime}$  both sectors are well
separated in energy, while at $K$ and $M$ they overlap. From $\Gamma$ to $M$ there is
a spectral weight  transfer from the high energy sector to the QP peak as it can be also
noticed quantitatively in the lower panel of Fig. \ref{fig2}.
In Fig. \ref{fig6} we show the magnetic
resonances above the QP peak for  $J/|t|=0.1$ ( at the ground state momentum $M$).
 We choose this particular parameter value in order to
make them more visible, although
they are clearly noticeable for $J/t\leq 0.5$ . In the inset it is shown
the energies of the first three resonances
{\it versus} $(J/t)^{2/3}$. The linear dependence of their energies allow us to associate
these resonances  with string
excitations of a hole confined by an effective linear potential of 
order $J$\cite{liu92,dagotto94}.
As we have previously stated,  there is a dynamical  enhancement of the  AF
environment around the hole when $t$ is negative, and this effect is
the responsible for the strong confining potential seen by the hole,
like in the square lattice case.

\begin{figure}[ht]
\vspace*{0.cm}
\includegraphics*[width=0.45\textwidth]{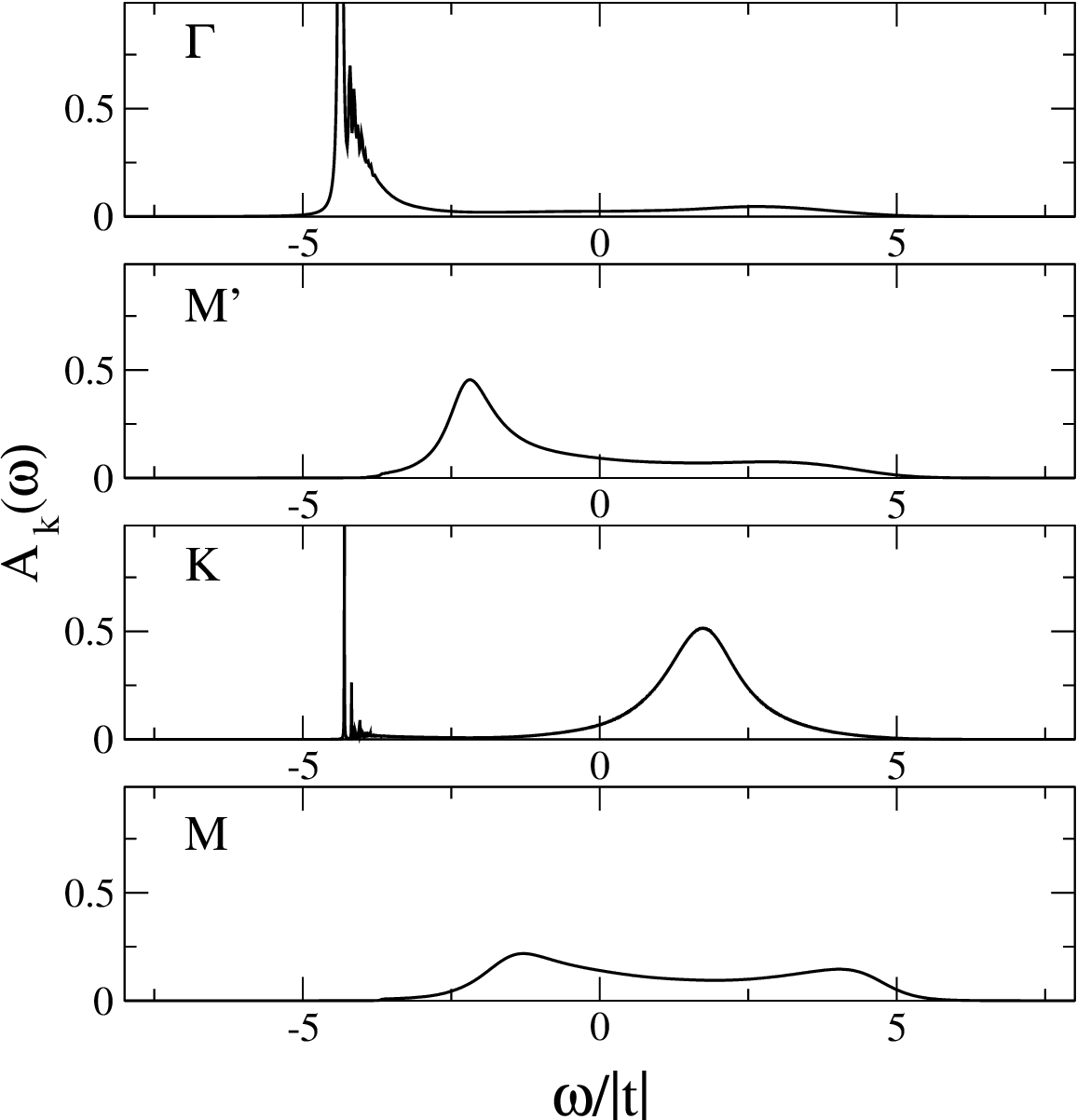}
\caption{Spectral function for the $\Gamma,M^{\prime},K,M$ points of the Brilluoin zone
for $J/|t|=0.4$ and positive $t$.}
\label{fig7}
\end{figure}

For positive $t$ (Fig. \ref{fig7}), the low energy sector consists of
a QP peak, when it exists ($\Gamma$ and $K$), and there is no 
signature of string-like excitations. 
This result is a manifestation of the ferromagnetic-like environment
around the hole,  pointed out before, for this $t$ sign.
Again the high energy sector has a broad bandwith of order $\sim9t$.
It is important to notice that at the ground state momentum
$\Gamma$ the QP is very robust, being the only relevant feature of the spectral function.
Otherwise, at $M^{\prime}$ and $M$ the spectral functions are almost structureless.

\section{Concluding remarks}

We have studied the effect of geometrical frustration on a single hole moving in
an antiferromagnetic background. We chose the triangular $t\!-\!J$ model
which at half filling presents a non-collinear magnetic ground state. Using the
spinless fermion representation and the self-consistent Born approximation we derived
an effective Hamiltonian with two different hole motion mechanisms. One of them
arises due to spin-flip processes (magnon-assisted) and the other one has its origin
from the ferromagnetic components of the non-collinear magnetic order. The latter allows
the free hole motion without emission or absorption of magnons and it is absent in
collinear antiferromagnets.
The reliability of the analytical method (SCBA) is supported by
the excellent agreement with the exact diagonalization method on small size clusters.

%*******MODIFICADO***************
Unlike previous works\cite{azzouz96,everts98},
we have performed a detailed description of the whole structures of the spectral function
and we were able to trace them back to the two hole motion processes
mentioned above. In particular, we have done a careful analysis of quasiparticles, strings,
and free hole resonances in the thermodynamic limit.

We have found new and reliable results of the low energy sector of the spectra.
Our main result is the vanishing of the quasiparticle excitation for
a {\it positive} integral  transfer $t$ at some momenta and at finite $J$.
Also, we have uncovered other important change of the spectra under $t$ sign
reversal: the presence of string-like excitations only for $t$ negative.
We suggest that these features are a consequence of different magnetic
environment around the hole tuned by the $t$ sign. Based on the quasiparticle energy
scaling with $J/t$,  we argue that for negative $t$
the local environment around the hole enhances its AF character, while for the other
sign the ferromagnetic character is favored. For small clusters
($N \le 12$), this behavior is reversed, in agreement with the three-sites problem
worked out by Koretsune and Ogata.\cite{koretsune02}
A deeper study of the magnetic
environment using the hole wave function within the SCBA will be presented
elsewhere \cite{trumper03}.

The dynamics of a hole in the triangular lattice, for negative $t$, exhibits the
same features of the well known square lattice case\cite{martinez91,liu92}:
a well defined quasiparticle, string excitations, and antiferromagnetic correlations
around the hole. On the contrary, for positive $t$ the dynamics is more irregular: 
no well defined quasiparticle, no string excitations, and ferromagnetic correlations
around the hole. 

For decades researchers have tried to look for spin-liquid
phases that upon hole doping, give rise to non-conventional excitations\cite{anderson87}.
Now we give firm evidence that such non-conventional excitations
can be found in {\it non-collinear} spin-crystal phases like the one
present in the triangular antiferromagnet.
%*****************************************************************************************

We acknowledge helpful discussions with A. E. Feiguin, A. Greco, A. Dobry, and
D. Poilblanc. A. E. T. thanks the hospitality received at the Universite Paul Sabatier in
Toulouse under the PICS France-Argentine 1490. We thank G. Martins for pointing out 
an error in an earlier version of this manuscript. 
This work was done under PICT Grant No. N03-03833 (ANPCyT) and was partially
supported by Fundaci\'on Antorchas.

\end{document}